\documentclass[aps,prd,preprint,superscriptaddress,amsmath,amssymb,showpacs]{revtex4-1}
\usepackage{dcolumn}
\usepackage{graphicx}
\usepackage{float}
\usepackage{physics}
\usepackage[colorlinks=true,allcolors=blue]{hyperref}
\begin{document}
	\title{Rotation effect on the deconfinement phase transition \\
		in holographic QCD
	}
	
	\author{Jia-Hao Wang}
	\affiliation{College of Science, China Three Gorges University, Yichang 443002, China}
	
	\author{Sheng-Qin Feng}
	\email{Corresponding author: fengsq@ctgu.edu.cn}
	\affiliation{College of Science, China Three Gorges University, Yichang 443002, China}
	\affiliation{Center for Astronomy and Space Sciences and Institute of Modern Physics, China Three Gorges University, Yichang 443002, China}
	\affiliation{Key Laboratory of Quark and Lepton Physics (MOE) and Institute of Particle Physics,\\
		Central China Normal University, Wuhan 430079, China}
	
	\date{\today}

	\begin{abstract}
		Abstract:  The impact of rotation on the deconfinement phase transition under the EM system of the soft and the hard wall models in holographic quantum chromodynamics is studied in this paper. The metric by cylindrical coordinates with rotation is introduced into the system to calculate the Hawking temperature. The first holographic study on the influence of the radius of a homogeneous rotating system on the phase diagram is proposed. It is found that the phase transition temperature hardly changes with the rotation angular velocity for a small rotation radius. Only with a larger rotation radius can the change in rotational angular velocity significantly alter the phase transition temperature. The phase transition temperature decreases rapidly with the increase of rotation angular velocity as the rotation radius increases.
	\end{abstract}
	
	\maketitle
	
	\section{Introduction}\label{sec:01_intro}

	The phase and properties of matter become very extraordinary under rotation, which has recently attracted a lot of interest among people. This type of research is particularly relevant to the strong interacting matter in quantum chromodynamics (QCD). For example, astrophysical objects composed of dense QCD matter such as neutron stars can spin rapidly \cite{RN1,RN2}. Typical noncentral nucleus-nucleus collisions can generate QCD matter that carries a nonzero angular momentum of the order of  $10^4$-$10^5 \hbar$ with local angular velocities in the range of 0.01-0.1 GeV \cite{RN3,RN4,RN5} in relativistic heavy ion collision experiments. An impressive progress \cite{RN6} has also achieved to study the rotating QCD matter by using lattice simulations.
	
	Exploring the effect of rotation on the phase transition of QCD matter is also very significant in relativistic heavy ion collisions. As is well known, noncentral nucleus- nucleus collisions can generate strong external magnetic fields \cite{RN7,RN8,RN9,RN10,RN11,RN12},  which have interesting effects on the thermodynamics and phase diagrams of QCD matter \cite{RN13,RN14,RN15,RN16,RN17,RN18}. The study of the magnetic catalytic and inverse magnetic catalytic properties of chiral condensates by using the NJL model \cite{RN19,RN20,RN21,RN61}, as well as the study of the effect of magnetic fields on phase diagrams by using holographic methods \cite{RN22,RN23,RN24,RN25,RN26,RN27,RN28,RN29,RN30} are currently in progressing smoothly. Given the close analogy between magnetic field and rotation, it is tempting to ask how rotation could affect phase transitions. In this article, we will use holographic methods to study the effect of rotation on QCD phase transition.
	
	It is well known that the confinement phase is at a lower temperature and density, while the deconfinement phase of QCD is at a higher temperature and density. The exploration of the phase structure of QCD is an important and challenging topic. How to probe the phase diagrams in the $T-\mu$ plane is a rather hard job because the QCD coupling constant becomes very large near the phase change region, and the traditional perturbation QCD method cannot be used. For a long time, the lattice QCD method is token as the only credible way to study the program. Although lattice QCD works well under zero baryon density, there is a sign problem when considering finite baryon density. However, the most interesting region in the QCD phase diagram is the region with finite baryon density. The QGP generated by heavy ion collisions, as well as compact stars in astrophysics, has a finite baryon density.
	
	In holographic QCD, various approaches exist to study the deconfinement phase transition. Among these, using the Hawking-Page phase transition \cite{RN31,RN32} from gravitational theory to investigate the QCD deconfinement phase transition proves to be highly effective. The primary physical idea involves the phase transition between thermal anti-de Sitter (AdS) at low temperature and the Schwarzschild AdS black hole at high temperature in pure gauge theory \cite{RN32}, which corresponds to the confinement-deconfinement phase transition between hadron state and QGP state in QCD. Reference \cite{RN33} delved into the Hawking-Page phase transition between the tAdS space and the Schwarzschild AdS black hole at zero chemical potential. Reference \cite{RN34,RN35,RN36} examined the Hawking-Page phase transition between thermal charged (tc) AdS space and the Reissner-Nordstr$\ddot{o}$m (RN) AdS black hole at nonzero chemical potential.
	
	In heavy ion collisions, in addition to the presence of a strong magnetic field, the plasma also exhibits significant angular momentum. There have been some studies that introduce rotation into holography \cite{RN37,RN38,RN39} and effective field theory \cite{RN40,RN41}. Braga \cite{RN37} utilized the corresponding Hawking-Page phase transition between the thermal AdS space and the Schwarzschild AdS black hole, and selected a rotating cylindrical symmetric black hole model to study the confinement-deconfinement phase transition and thermodynamics of a rotating system with zero chemical potential. In this article, we will expand corresponding Hawking-Page phase transition into between thermal charged anti-de Sitter (AdS) space and RNAdS black hole, discussing the confinement-deconfinement phase transition and thermodynamic properties of nonzero chemical potentials under rotational conditions.
	
	The purpose of this article is to investigate the effect of plasma rotation on the phase transition temperature $T_c$. This article will use holographic AdS/QCD models, especially hard wall \cite{RN42,RN43,RN44} and soft wall \cite{RN16,RN45} models, to discuss the dependence of phase transition temperature $T_c$ on rotational angular velocity, rotational radius and chemical potential. The structure of the paper is as follows: In Sec. II, we introduce the rotation into holographic QCD and discuss the consequent impacts on temperature and chemical potential. Section III delves into the effects of rotation on the deconfinement phase transition by utilizing both the hard and soft wall models. A summary and discussion are presented in Sec. IV.
	
	\section{The holographic system under rotational background}\label{sec:02 setup}
	
	In this section, we will introduce some common theoretical settings in the context of rotation. To analyze the rotating charged black hole, it is necessary to assume that the spacetime has cylindrical symmetry \cite{RN38,RN39,RN46,RN47,RN48}. Consequently, the metrics of Lorentzian signature in the cylindrical form of both AdS black hole (bh) and thermal charged (tc) AdS space can be uniformly reformulated as
\begin{equation}
	d s _ { b h , t c } ^ { 2 } = \frac { L ^ { 2 } } { z ^ { 2 } } ( - f _ { b h , t c } ( z ) d t ^ { 2 } + d \vec{ x } ^ { 2 } + l ^ { 2 } d \phi ^ { 2 } + \frac { 1 } { f _ {  b h , t c } ( z ) } d z ^ { 2 } ) ,
\end{equation}
where $L$ is AdS radius, $\phi$ is the angular coordinate describing the rotation, and $l$ is the radius of the rotation axis. The metric function of Eq. (1) for the background geometry of AdS black hole is given as
\begin{equation}
	f _ { b h } ( z ) = 1 - m z ^ { 4 } + q ^ { 2 } z ^ { 6 } ,
\end{equation}
where $m$ and $q$ are the black hole mass and charge, respectively.

The background geometry of thermal charged AdS space is
\begin{equation}
	f _ { t c } ( z ) = 1 + q _ { 1 } ^ { 2 } z ^ { 6 } ,
\end{equation}
where $q_1$ is the charge of thermal charged AdS space. Because of boundary condition $f_{bh}(z_h)=0$, one can obtain the mass as
\begin{equation}
	m = \frac { 1 } { z _ { h } ^ { 4 } } + q ^ { 2 } z _ { h } ^ { 2 } .
\end{equation}

Next, the rotation will be introduced into the QCD medium system. According to Ref. \cite{RN49}, one obtains the rotating extension from the static configuration through a local Lorentz boost as
\begin{equation}
	t \rightarrow \frac { 1 } { \sqrt { 1 - l ^ { 2 } \omega ^ { 2 } } } ( t + l ^ { 2 } \omega \phi ) ,
\end{equation}
and
\begin{equation}
	\phi \rightarrow \frac { 1 } { \sqrt { 1 - l ^ { 2 } \omega ^ { 2 } } } ( \phi + \omega t ) ,
\end{equation}
where $\omega$ is the angular velocity. The corresponding transformation of the metric is given as
\begin{equation}
	d s ^ { 2 } = g _ { tt } d t ^ { 2 } + g _ { t\phi  } d t d \phi + g _ { \phi  t } d \phi d t + g _ { \phi \phi } l ^ { 2 } d \phi ^ { 2 } + g _ { z z } d z ^ { 2 } + g _ { x x } \sum _ { i = 1 } ^ { 2 } d x _ { i } ^ { 2 } ,
\end{equation}
with
\begin{equation}
	g _ { tt } = \frac { \gamma ^ { 2 } ( \omega l ) L ^ { 2 } } { z ^ { 2 } } ( \omega ^ { 2 } l ^ { 2 } - f ( z ) ) ,
\end{equation}
\begin{equation}
	g _ { \phi \phi } = \frac { \gamma ^ { 2 } ( \omega l ) L ^ { 2 } } { z ^ { 2 } } ( 1 - \omega ^ { 2 } l ^ { 2 } f ( z ) ) ,
\end{equation}
\begin{equation}
	g _ { t \phi } = g _ { \phi t } = \frac { \gamma ^ { 2 } ( \omega  l ) L ^ { 2 } } { z ^ { 2 } } \omega l ^ { 2 } ( 1 - f ( z ) ) ,
\end{equation}
\begin{equation}
	g _ { zz } = \frac { L ^ { 2 } } { z ^ { 2 } f ( z ) } ,
\end{equation}
\begin{equation}
	g _ { x x } = \frac { L ^ { 2 } } { z ^ { 2 } } ,
\end{equation}
where $\gamma(\omega l)$ is the Lorentz factor
\begin{equation}
	\gamma ( \omega l ) = \frac { 1 } { \sqrt { 1 - l ^ { 2 } \omega ^ { 2 } } } .
\end{equation}

The Lorentz boost determines the relation between the Hawking temperature $T$ of the black hole and the horizon position. Reestablishing the canonical form of rotation metric given by Eq. (7), as shown in Refs. \cite{RN38,RN46}, one yields the following metric as
\begin{equation}
	d s ^ { 2 } = \frac { L ^ { 2 } } { z ^ { 2 } } \left[ - N ( z ) ^ { 2 } f ( z ) d t ^ { 2 } + \frac { d z ^ { 2 } } { f ( z ) } + R ( z ) ( d \phi + P ( z ) d t ) ^ { 2 } + \sum _ { i = 1 } ^ { 2 } d x _ { i } ^ { 2 } \right] ,
\end{equation}
with
\begin{equation}
	N ( z ) ^ { 2 } = \frac { ( 1 - \omega ^ { 2 } l ^ { 2 } ) } { 1 - f ( z ) \omega ^ { 2 } l ^ { 2 } } ,
\end{equation}
\begin{equation}
	R ( z ) = \gamma ^ { 2 } l ^ { 2 } - f ( z ) \gamma ^ { 2 } \omega ^ { 2 } l ^ { 4 } ,
\end{equation}
\begin{equation}
	P ( z ) = \frac { \omega ( 1 - f ( z ) ) } { 1 - f ( z ) \omega ^ { 2 } l ^ { 2 } } ,
\end{equation}
where $N(z)$ is the lapse function and $P(z)$ is the shift function.

Following Refs. \cite{RN49,RN50}, one can obtain the following expressions for the Hawking temperature of rotating black holes:
\begin{equation}
	T _ { H } = - \frac { N ( z _ { h } ) f ^ { \prime } ( z _ { h } ) } { 4 \pi } .
\end{equation}

It should be noted that due to the Lorentz transformation of time and angular rotation components, the expression of the chemical potential differs from that of the static case. The quark chemical potential under rotation can be given as \cite{RN51,RN52}
\begin{equation}
	\mu = A _ { \mu } \chi ^ { \mu } | _ { z = z _ { h } } - A _ { \mu } \chi ^ { \mu } | _ { z = 0 } ,
\end{equation}
where $\chi = \partial _ { t } + \Omega \partial _ { \phi } $ is the Killing vector, and the gauge potential $A_\mu$ under rotation is given as
\begin{equation}
	A _ { \mu } = A _ { t } ( \gamma \delta _ { \mu } ^ { t } + \gamma \omega l ^ { 2 } \delta _ { \mu } ^ { \phi } ) .
\end{equation}

Substituting (20) into (19), one obtains the chemical potential as
\begin{equation}
	\mu = \mu ^ { \prime } \sqrt { 1 - \omega ^ { 2 } } .
\end{equation}

The Ricci tensor $R_{mn}$ and metric $g_{mn}$ are second rank tensors, which transform under coordinate transformation of rotation to
\begin{equation}
	T _ { \alpha \beta } ^ { \prime } = \frac { \partial x ^ { m } } { \partial x ^ { \prime \alpha } } \frac { \partial x ^ { n } } { \partial x ^ { \prime \beta } } T _ { m n } \equiv M _ { \alpha \beta } ^ { m n } T _ { m n } .
\end{equation}

However, for scalar coordinate transformations, the situation is different. For example: as one defines $\mathcal{R}$ as the curvature scalar, and $\mathcal{R}^\prime$ as the curvature scalar with coordinate transformation of rotation, this transformation relation should be
\begin{equation}
	\mathcal{R} ^ { \prime } = g ^ { m n } R _ { m n } = M _ { \alpha \beta } ^ { m n } g ^ { \alpha \beta } M _ { m n } ^ { \alpha \beta } R _ { \alpha \beta } = g ^ { \alpha \beta } R _ { \alpha \beta } = \mathcal{R} ,
\end{equation}
so, we can conclude that scalar quantity is not affected by the coordinate transformation of rotation.

\section{The phase diagram under rotation background}\label{sec:03 setup}

In this section, we will discuss holographic QCD phase transition using the hard wall and soft wall model, respectively.

\subsection{Hard wall model}

Following Refs. \cite{RN34,RN36}, the Euclidean action of the Einstein-Maxwell (EM) system describing holographic light quarks for the hard wall model is given by:
\begin{equation}
	S = \int d ^ { 5 } x \sqrt { G } \left[ \frac { 1 } { 2 \kappa  ^ { 2 } } ( - \mathcal{R} + 2 \Lambda ) + \frac { 1 } { 4 g ^ { 2 } } F _ { M N } F ^ { M N } \right] ,
\end{equation}
where $G$ is the determinant of metric, $\kappa$ is the five-dimensional Newton constant, $\mathcal{R}$ is the Ricci scalar, $\Lambda$ is the cosmological constant, $g$ is the 5D gauge coupling constant, and $F_{MN}$ is the U(1) bulk gauge field strength tensor with $F_{MN} = \partial_M A_N-\partial_N A_M$. According to Refs. \cite{RN34,RN53,RN54}, the bulk gauge field $A_M$ is
\begin{equation}
	\begin{aligned}
		A _ { t } = \frac { \mu } { \sqrt { 1 - \omega ^ { 2 } l ^ { 2 } } } - \rho z ^ { 2 } \ ,\  	A _ { i } = A _ { z } = 0 \ ( i = 1 , 2 , 3 ),
	\end{aligned}
\end{equation}
where $\rho$ is the quark number density, which is related to the black hole charge $q$
\begin{equation}
	\rho = \sqrt { \frac { 3 g ^ { 2 } L ^ { 2 } } { 2 k ^ { 2 } } } q .
\end{equation}

By imposing the Dirichlet boundary condition $A_t(z_h)=0$ near the boundary at horizon, one can rewrite $\rho$ as
\begin{equation}
	\rho = \frac { \mu } { z _ { h } ^ { 2 } \sqrt { 1 - \omega ^ { 2 } l ^ { 2 } } } .
\end{equation}

The five-dimensional Newton constant $\kappa$, and the 5D gauge coupling constant $g$ are given as \cite{RN54},
\begin{equation}
	\frac { 1 } { \kappa  ^ { 2 } } = \frac { N _ { c } ^ { 2 } } { 4 \pi ^ { 2 } L ^ { 3 } } ,
\end{equation}
and
\begin{equation}
	\frac { 1 } { g ^ { 2 } } = \frac { N _ { c } N _ { f } } { 4 \pi ^ { 2 } L } ,
\end{equation}
where $N_c$ and $N_f$ are the number of colors and flavors. From Eq. (26) to Eq. (29), one can obtain
\begin{equation}
	q = \sqrt { \frac { 2 N _ { f } } { 3 N _ { c } } } \frac { \mu } { z _ { h } ^ { 2 } \sqrt { 1 - \omega ^ { 2 } l ^ { 2 } } } .
\end{equation}

The boundary condition at the IR cut-off $z = z_{IR}$ is
\begin{equation}
	A ( z _ { I R } ) = \alpha \frac { \mu } { \sqrt { 1 - \omega ^ { 2 } l ^ { 2 } } } ,
\end{equation}
where $\alpha$ is a parameter, which can be determined as $\alpha = -\frac{1}{2}$ given by Ref. \cite{RN34}. The quark number density $\rho$ is given
\begin{equation}
	\rho = \frac { 3 \mu } { 2 z _ {I R } ^ { 2 } \sqrt { 1 - \omega ^ { 2 } l ^ { 2 } } } .
\end{equation}

By comparing Eq. (25) and Eq. (31), one can derive the charge of thermal charged AdS space $q_1$ as
\begin{equation}
	q _ { 1 } = \sqrt { \frac { 3 N _ { f } } { 2 N _ { c } } } \frac { \mu } { z _ { IR } ^ { 2 } \sqrt { 1 - \omega ^ { 2 } l ^ { 2 } } } .
\end{equation}

From Eq. (24), the corresponding on-shell gravity action of RNAdS BH and tcAdS space for the hard wall model are
\begin{equation}
	S _ { B H } = \frac { L ^ { 3 } V _ { 3 } } { \kappa ^ { 2 } } \int _ { 0 } ^ { \beta } dt \int _ { \epsilon } ^ { z _ { h } } d z ( \frac { 4 } { z ^ { 5 } } - 2 q ^ { 2 } z ) ,
\end{equation}
\begin{equation}
	S _ { t c } = \frac { L ^ { 3 } V _ { 3 } } { \kappa  ^ { 2 } } \int _ { 0 } ^ { \beta^\prime } d t \int _ { \epsilon  } ^ { z _ { IR } } d z ( \frac { 4 } { z ^ { 5 } } - 2 q _ { 1 } ^ { 2 } z ) ,
\end{equation}
where $V_3$ is the volume of the 3D space, $\epsilon(\epsilon\to0)$ is the UV cut-off, $\beta$ and $\beta^\prime$ are the time periodicity in RNAdS BH and tcAdS space, respectively. At $z=\epsilon$, $\beta^\prime$ is related to $\beta$ as follows:
\begin{equation}
	\beta ^ { \prime } \sqrt { f _ { t } ( \epsilon ) } = \beta \sqrt { f _ { b } ( \epsilon ) } .
\end{equation}

The on-shell action density difference between RNAdS BH and tcAdS space is given as:
\begin{equation}
	\Delta {\text{\LARGE$\varepsilon$}} = \frac { S _ { B H } - S _ { t c } } { V _ { 3 } } = \frac { L ^ { 3 } } { \kappa  ^ { 2 } } \left[ \beta \int _ { \epsilon  } ^ { z _ { h } } d z \left( \frac { 4 } { z ^ { 5 } } - 2 q ^ { 2 } z \right) - \beta ^ { \prime } \int _ { \epsilon  } ^ { z _{ IR } } d z \left( \frac { 4 } { z ^ { 5 } } - 2 q _ { 1 } ^ { 2 } z \right) \right].
\end{equation}

The free energy density is given as $F = - ( T \ln \mathcal{Z} ) / V _ { 3 } $, where $\mathcal{Z}$ is the partition function. It should be noted that, in our pursuit of studying the thermodynamics of this hard wall model, we utilize the compactified imaginary time metric (Euclidean signature). Thus, the free energy is expressed as:
\begin{equation}
	F = - \frac { T \ln \mathcal{Z} } { V _ { 3 } } \approx \frac { T S _ { E } } { V _ { 3 } } .
\end{equation}
	
	Therefore, the difference of free energy density between RNAdS BH and tcAdS space is
	\begin{equation}
		\begin{aligned}
			\Delta F =& T \Delta {\text{\LARGE$\varepsilon$}} =  \frac { 3 N _ { c } ( - 1 + \omega ^ { 2 } l ^ { 2 } ) + N _ { f } z _ { h } ^ { 2 } \mu ^ { 2 } } { 1 8 N _ { c } ^ { 2 } \pi ( 1 - \omega ^ { 2 } l ^ { 2 } ) ^ { \frac { 3 } { 2 } } z _ {I R } ^ { 4 } z _ { h } ^ { 5 } }\\
			&\times[ - 3 N _ { c } ( - 1 + \omega ^ { 2 } l ^ { 2 } ) ( z _ { I R } ^ { 4 } - 2 z _ { h } ^ { 4 } ) + N _ { f } z _ { I R } ^ { 2 } z _ { h } ^ { 2 } ( 2 z _ { I R } ^ { 2 } - 9 z _ { h } ^ { 2 } ) \mu ^ { 2 }],
		\end{aligned}
	\end{equation}
	where $z_{IR} = 1/(0.323 GeV)$ is obtained from the lightest $\rho$ meson mass \cite{RN33}.
	
	The dependence of $\Delta F$ on $z_h$ with different $\mu$ and $\omega$ for the hard wall with $m = N_f/N_c =1$ is manifested in Fig. 1.
	
	When $\Delta F$ is positive, the tcAdS space is stable. On the contrary, when $\Delta F$ is negative, the RNAdS black hole is stable. According to Hawking-Page phase transition, the tcAdS space corresponds to confinement phase, the RNAdS black hole corresponds to deconfinement phase. Therefore, a positive (negative) $\Delta F$ suggests that the QCD system is in the confinement (deconfinement) phase. Thus, when the difference of free energy density equals zero, the corresponding horizon position marks the phase transition point for the Hawking-Page phase transition or confinement-deconfinement phase transition.
	\begin{figure}[H]
		\centering
		\includegraphics[width=1\textwidth]{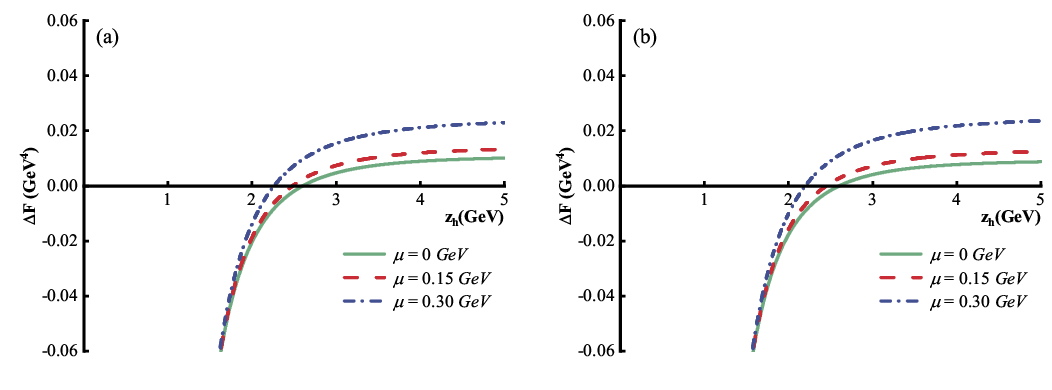}
		\caption{\label{fig1} The difference of free energy density of rotational system $\Delta F$ as a function of the horizon position $z_h$ for different chemical potential for the hard wall model. (a) with $\omega$ = 0 GeV. and (b) with $\omega$ = 0.5 GeV}
	\end{figure}
	
	From Fig. 1(a) and 1(b), one finds that at a constant angular velocity $\omega$, the horizon position of the phase transition point becomes small with the increase of the chemical potential $\mu$. Comparing Fig. 1(a) and Fig. 1(b), one also finds that at a fixed $\mu$, the horizon position of the phase transition point becomes small with the increase of the angular velocity $\omega$. It is found that chemical potential and angular velocity have the same effects on the horizon transition position., both of which cause the position of the phase transition point to decrease. Let us figure out what cause the reason in the following.
	
	In the context of a rotating QCD system, we know that the introduction of rotation alters various physical quantities. Refs. \cite{RN55,RN56} point out that the influence of rotation on thermodynamic quantities should have a modification of the thermodynamic relation as
	\begin{equation}
		\epsilon = - P + T s + \mu \rho + \omega J ,
	\end{equation}
	where $\epsilon$, $P$, $s$, $\rho$, and $J$ are the energy density, pressure, entropy density, quark number density, and angular momentum. In (40), it is evident that the energy density of the system also increases with $\omega$ and $\mu$ increase. Following the law of the black hole thermodynamics \cite{RN57}, the energy density of black hole correlates with its other properties as follows:
	\begin{equation}
		d E \propto d M \propto d A \propto d r \propto \frac { 1 } { z _ { h } } ,
	\end{equation}
	where $M$, $A$, and $r$ denote the mass, area and radius of black hole, respectively. One can find that the horizon position decreases with the increase of energy density. This explains why $z_h$ decreases with the increase of $\omega$ and $\mu$.
	
	In order to describe the status of a two-dimensional ideal rotating system, two critical physical quantities must be considered: angular velocity and radii of rotation. Therefore, our study encompasses not only the variation of phase transitions with rotation and chemical potential but also its phase diagram on the radius of rotation.
	
	In Fig. 2, the phase transition temperature is shown as a function of the angular velocity with different chemical potential for different rotational radiuses $l$ = 0.3GeV$^{-1}$, $l$ = 0.6GeV$^{-1}$ and $l$ = 0.9GeV$^{-1}$, respectively. One finds that small angular velocity has less impact on transition temperature while large angular velocity induces a quick decrease of phase transition temperature. It is also found that the critical temperature $T_c$ decreases with the increase of angular velocity but has weak dependence of chemical potential, which is analogous to some holographic work and PNJL model of deconfinement transition of light flavor \cite{RN58,RN59,RN60}. There it has been found that the presence of anisotropy of rotation causes easier dissociation of the $Q\bar Q$ and that in phase transitions anisotropy takes a catalytic reduction of the critical temperature.
	
	Since we are discussing the rotational system of QCD medium, the phase transition feature should rely on the finite size of the rotational system. Due to the cylindrical symmetry, these quantities are dependent on the rotation radius $l$. It would be interesting to probe how the phase transition feature in a strongly interacting rotating matter depends on the radius of the rotating system. The properties as a function of the radius of the rotating system may be connected to experimental observations in the future. And due to the limitation of the speed of light, it naturally causes the limitations $\omega l \leq 1$. Comparing Fig. 2(a) ($l$ = 0.3GeV$^{-1}$), Fig. 2(b) ($l$ = 0.6GeV$^{-1}$) and Fig. 2 is the(c) ($l$ = 0.9GeV$^{-1}$),  one finds that the phase transition temperature hardly changes with the rotation angular velocity for small rotation radii (l = 0.3GeV$^{-1}$), but as the rotation radius increases, the phase transition temperature decreases rapidly with the increase of rotation angular velocity.
	\begin{figure}[H]
		\centering
		\includegraphics[width=0.35\textwidth]{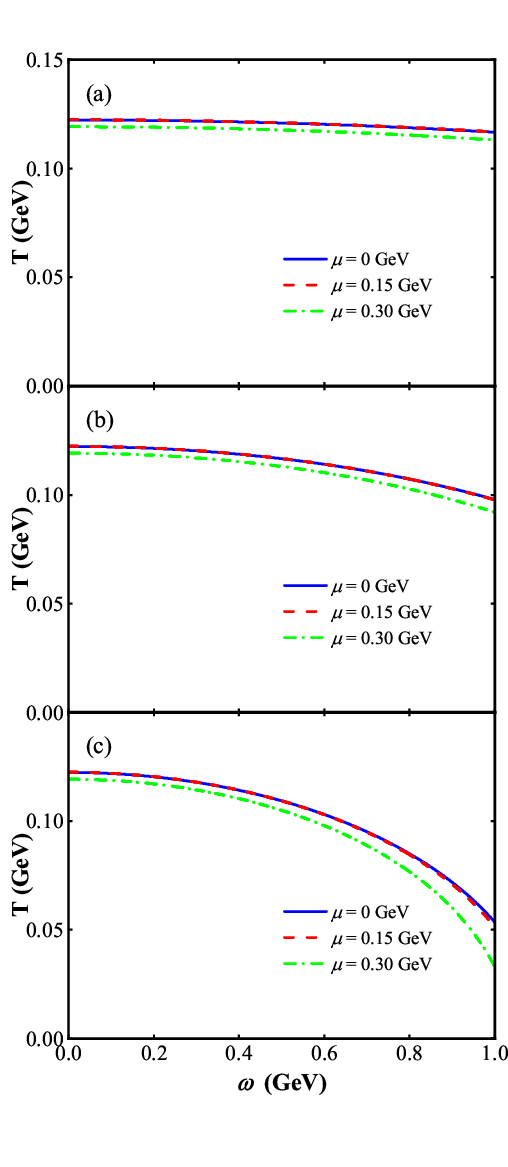}
		\caption{\label{fig2} Phase diagrams $T-\omega $ with different chemical potential of $\mu$ = 0 GeV, $\mu$ = 0.15 GeV and $\mu$ = 0.3 GeV for the hard wall model. (a) for $l$ = 0.3 GeV$^{-1}$, (b) for $l$ = 0.6 GeV$^{-1}$ and (c) for $l$ = 0.9 GeV$^{-1}$. }
	\end{figure}
	
	\subsection{Soft wall model}
	
	For soft wall model \cite{RN35}, the Euclidean action of the EM system is given by:
	\begin{equation}
		S = \int d ^ { 5 } x \sqrt { G } e ^ { - \Phi } \left[ \frac { 1 } { 2 \kappa  ^ { 2 } } ( - \mathcal{R}  + 2 \Lambda ) + \frac { 1 } { 4 g ^ { 2 } } F _ { M N } F ^ { M N } \right] ,
	\end{equation}
	where $\Phi$ is the nondynamical scalar field with $\Phi = c z^2$, $c$ is the IR energy parameter. The RNAdS BH charge $q$ and tcAdS space charge $q_1$ under the rotation background in the soft wall model are given as
	\begin{equation}
		q = \sqrt { \frac { 2 N _ { f } } { 3 N _ { c } } } \frac { \mu } { z _ { h } ^ { 2 } \sqrt { 1 - \omega ^ { 2 } } } ,
	\end{equation}
and
	\begin{equation}
		q _ { 1 } = \sqrt { \frac { 3 N _ { f } } { 2 N _ { c } } } \frac { \mu c } { \sqrt { 1 - \omega ^ { 2 } } } .
	\end{equation}
	
	The corresponding on-shell gravity action of RNAdS BH and tcAdS space are
	\begin{equation}
		S _ { B H } = \frac { L ^ { 3 } V _ { 3 } } { \kappa  ^ { 2 } } \int _ { 0 } ^ { \beta } d t \int _ { \epsilon  } ^ { z _ { h } } d z e ^ { - c z ^ { 2 } } ( \frac { 4 } { z ^ { 5 } } - 2 q ^ { 2 } z ) ,
	\end{equation}
	\begin{equation}
		S _ { t c } = \frac { L ^ { 3 } V _ { 3 } } { \kappa  ^ { 2 } } \int _ { 0 } ^ { \beta^\prime  } dt \int _ { \epsilon  } ^ { \infty } d z e ^ { - cz ^ { 2 } } ( \frac { 4 } { z ^ { 5 } } - 2 q _ { 1 } ^ { 2 } z ) .
	\end{equation}
	
	Contrasted with the hard wall model, the integral term is multiplied by a quadratic exponent of $z$, and the upper limit of the integral extends from $z_{IR}$ to infinity for the soft wall model. The free energy density difference $\Delta F$ between RNAdS BH and tcAdS space is given as
	\begin{equation}
		\begin{aligned}
			\Delta F =& T \Delta {\text{\LARGE$\varepsilon$}} = - \frac { 1 } { 6 c N _ { c } \sqrt { 1 - \omega ^ { 2 } l ^ { 2 } } z _ { h } ^ { 4 } } e ^ { - c z _ { h } ^ { 2 } } \\
			          &\times \{(6 c ^ { 3 } N _ { c } z _ { h } ^ { 4 } e ^ { c z _ { h } ^ { 2 } } ( \omega ^ { 2 } l ^ { 2 } - 1 ) E i ( - c z _ { h } ^ { 2 } ) - 3 c ^ { 2 } z _ { h } ^ { 2 } ( 3 \mu ^ { 2 } N _ { f } z _ { h } ^ { 2 } e ^ { c z_ { h } ^ { 2 } } + N _ { c } ( 2 - 2 \omega ^ { 2 } l ^ { 2 } ) )\\
			          &+ c ( 3 N _ { c } ( e ^ { c z_ { h } ^ { 2 } } - 2 ) ( \omega ^ { 2 } l ^ { 2 } - 1 ) - 2 \mu ^ { 2 } N _ { f } z _ { h } ^ { 2 } e ^ { c z_ { h } ^ { 2 } } ) + 4 \mu ^ { 2 } N _ { f } ( e ^ { c z_ { h } ^ { 2 } } - 1 )\} ,
		\end{aligned}
	\end{equation}
	where the parameter $c$ can be matched to the mass of the lightest $\rho$ meson \cite{RN33}, denoted as $\sqrt{c}$ = 0.338GeV, and $E_i(z)$ is the exponential integral function, which can be expressed as
	\begin{equation}
		E i ( z ) = - \int _ { - z } ^ { \infty } d t \frac { e ^ { - t } } { t } .
	\end{equation}
	\begin{figure}[H]
		\centering
		\includegraphics[width=0.35\textwidth]{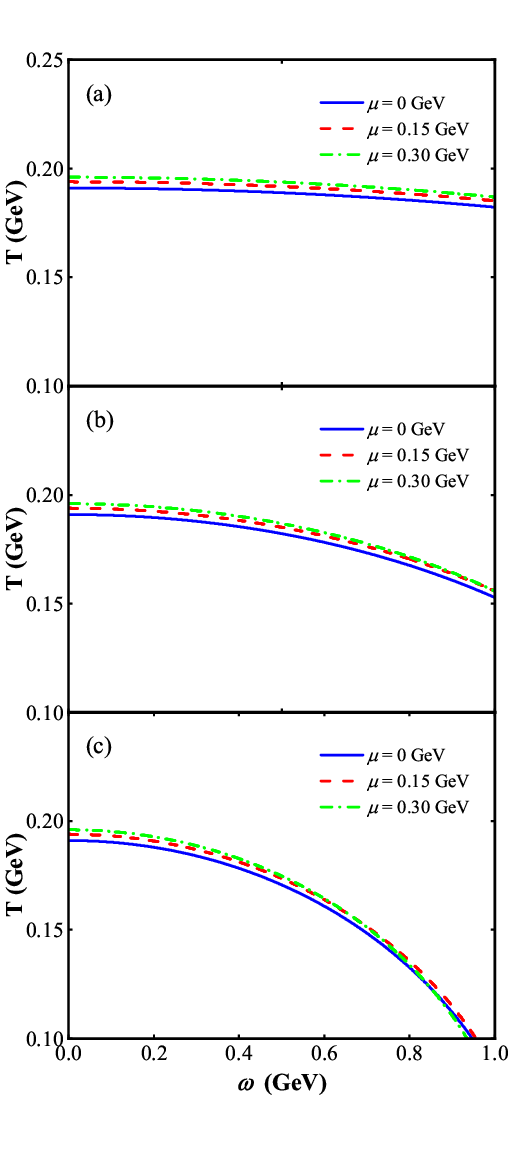}
		\caption{\label{fig3} Phase diagrams $T-\omega$ with different chemical potential of $\mu$ = 0 GeV, $\mu$ = 0.15 GeV and $\mu$ = 0.3 GeV for the soft wall model. (a) for $l$ = 0.3 GeV$^{-1}$, (b) for $l$ = 0.6 GeV$^{-1}$ and (c) for $l$ = 0.9 GeV$^{-1}$.}
	\end{figure}
	Figure 3 shows the phase diagrams $T-\omega$ with different chemical potentials and different rotation radii for the soft wall model. It is found that the characteristics of the phase diagram of soft wall case are generally similar to that of hard wall case, but there are also some differences. For example, when analyzing the dependence of phase transition temperature on chemical potential, it is found that for the hard wall model, the larger the chemical potential is, the smaller the phase transition temperature is. However, for the soft wall model, the larger the chemical potential is, the larger the phase transition temperature is.
	
	\subsection{The comparison of phase diagram between the hard wall and the soft wall}
	
	The comparisons of phase diagram between the hard wall and the soft wall are published in Fig.4. Compared the soft wall model with hard wall model, the trend of phase transition temperature $T_c$ decreasing with angular velocity $\omega$ is almost the same. However, the phase transition temperature given by soft wall model is obviously larger than that of hard wall model under the same conditions. As the rotation radius $l$ increases, the decreasing speed of the phase transition with angular velocity for the soft wall model is faster than that of the hard wall model.
	\begin{figure}[H]
		\centering
		\includegraphics[width=0.35\textwidth]{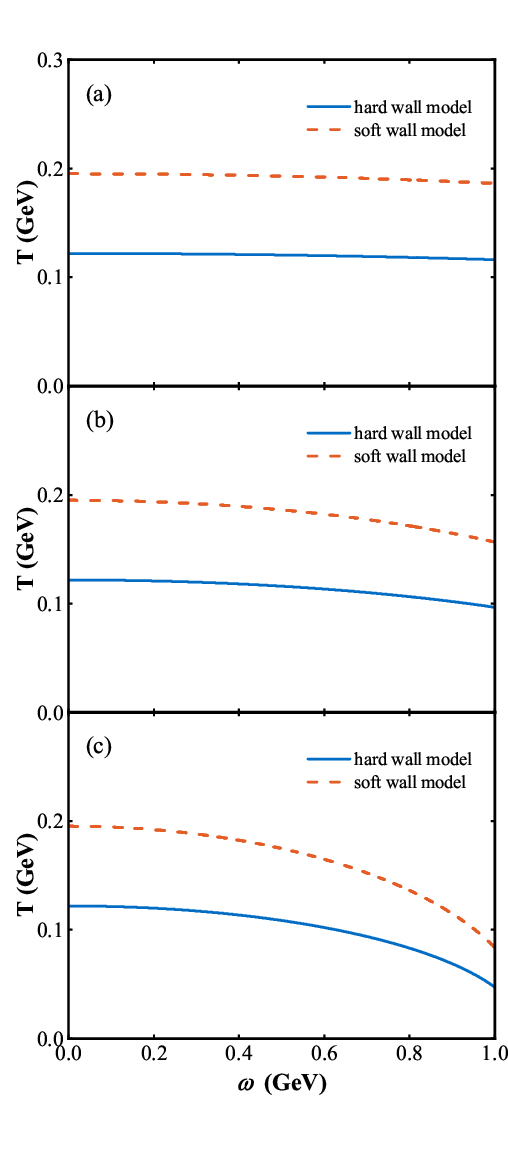}
		\caption{\label{fig4} The phase diagram $T-\omega$ comparison between the hard wall model and the soft wall model with chemical potential of $\mu$ = 0.2 GeV. (a) for $l$ = 0.3 GeV$^{-1}$, (b) for $l$ = 0.6 GeV$^{-1}$ and (c) for $l$ = 0.9 GeV$^{-1}$.}
	\end{figure}
	
	\section{SUMMARY AND CONCLUSIONS}\label{sec:04 summary}
	
	In this paper, we focus on the impact of rotation on the deconfinement phase transition under the EM system of the soft and the hard wall model in holographic QCD. The metric by cylindrical coordinates with rotation is introduced into the system to calculate the Hawking temperature. Whether it is a soft wall model or a hard wall model, the phase transition temperatures decrease with the increase of rotation angular velocity, but have weak dependence of chemical potential.
	
	The first holographic study on the influence of the radius of a homogeneous rotating system on the phase diagram is established in the article. As we are discussing the rotating system of QCD medium, the phase transition characteristics should depend on the finite size of the rotating system. Due to the cylindrical symmetry of the rotating system, the rotation radius $l$ has become an important characteristic quantity of the rotating system. Studying the dependence of phase transition characteristics of the strongly interacting rotating matter on the radius of the rotating system is an important research topic.  It is found that the phase transition temperature hardly changes with the rotation angular velocity for a small rotation radius, but as the rotation radius increases, the phase transition temperature decreases rapidly with the increase of rotation angular velocity.
	
	\section*{Acknowledgments}
	This work was supported by the National Natural Science Foundation of China (Grants No. 11875178, No. 11475068, No. 11747115).
	
	\section*{References}
	
	\nocite{*}
	\bibliography{ref}
	
\end{document}